\documentclass[Times,4pt,aps,pra,twocolumn,showpacs,amsmath,amssymb,floatfix,footinbib,superscriptaddress]{revtex4-2}
\usepackage{amsmath,natbib,empheq,graphicx,subfigure,amssymb,graphics,amsmath,mathrsfs,CJK,color,comment}
\usepackage{multirow,fancyhdr,color,bm,tabularx,psfrag,geometry,dcolumn,datetime,physics,booktabs}
\usepackage[colorlinks=true,linkcolor=blue,urlcolor=blue,citecolor=blue]{hyperref}
\usepackage[mathlines]{lineno}
\def\footnoterule{\kern -1mm \hrule width 5.8cm \kern 2.2mm}
\usepackage{tikz,xcolor,hyperref}
\definecolor{lime}{HTML}{A6CE39}
\DeclareRobustCommand{\orcidicon}{%
    \begin{tikzpicture}
    \draw[lime, fill=lime] (0,0)
    circle [radius=0.16]
    node[white] {{\fontfamily{qag}\selectfont \tiny ID}};\draw[white, fill=white] (-0.0625,0.095)
    circle [radius=0.007];
    \end{tikzpicture}
    \hspace{-2mm}}
\foreach \x in {A, ..., Z}
{\expandafter\xdef\csname orcid\x\endcsname{\noexpand\href{https://orcid.org/\csname orcidauthor\x\endcsname}{\noexpand\orcidicon}}}

\begin{document}

\title{Catalytic Stabilization of Ergotropy and Backflow Suppression in Open Many-Body Quantum Batteries}%

\author{Liang Luo}
\affiliation{Center for Quantum Materials and Computational Condensed Matter Physics, Faculty of Science, Kunming University of Science and Technology, Kunming, 650500, PR China}
\author{Shun-Cai Zhao\orcidA{}}
\email[Corresponding author: ]{zsczhao@126.com.}
\affiliation{Center for Quantum Materials and Computational Condensed Matter Physics, Faculty of Science, Kunming University of Science and Technology, Kunming, 650500, PR China}
\date{\currenttime,~\today}

\begin{abstract}
Coherent energy backflow and non-Markovian oscillations limit energy retention and degrade extractable work (ergotropy) in open many-body quantum batteries. Here, we present a catalyst-mediated charging protocol for a collective spin-array quantum battery coupled to a laser-driven charger. Using the open-system Lindblad master equation, we examine the energy transfer dynamics when both charger and battery are symmetrically coupled to an off-resonant auxiliary catalytic mode. Numerical simulations reveal that while unassisted bipartite setups exhibit pronounced backflow oscillations and poor energy retention, catalytic mediation quenches transient oscillations and accelerates energy injection. The auxiliary system operates as an energy-invariant conduit, maintaining a constant energy expectation value $\langle H_C(t)\rangle \approx \langle H_C(0)\rangle$ and negligible transient population throughout the evolution. Microscopically, virtual excitations of the catalyst generate an effective complex inter-subsystem coupling $J_{\text{eff}}$, which induces an underdamped-to-overdamped dynamical crossover and introduces selective coherence damping. This mechanism prevents population depletion in the battery, stabilizing the population inversion and significantly increasing the asymptotic steady-state ergotropy with increasing battery size $N_B$. These findings clarify the dissipative dynamics of catalyst-mediated energy transfer and provide a practical scheme for improving storage stability in modern quantum hardware platforms.
\end{abstract}
\keywords{ Many-Body Quantum Batteries；ergotropy；energy backflow； catalyst-mediated charging}
\maketitle
\section{Introduction}

Operating at the intersection of quantum information science and nonequilibrium quantum thermodynamics, quantum batteries (QBs) serve as a central paradigm for energy storage at the microscopic scale \cite{Preskill2018quantumcomputingin,Goold2016,Vinjanampathy01102016}. By exploiting genuine quantum resources such as coherence \cite{Caravelli_2021}, multipartite entanglement \cite{PhysRevLett.129.130602}, and collective cooperative phenomena, QBs can achieve charging powers and capacities that surpass classical electrochemical limits \cite{Hymas2026, Hotta2025}. Since the foundational theoretical work by Alicki and Fannes \cite{PhysRevE.87.042123}, research has expanded from elementary few-qubit architectures \cite{Yu2023} to complex many-body setups \cite{3-xqtv-qbyk}, establishing fundamental upper bounds on charging speeds \cite{Juli_Farr__2020, Puri2024} with prospective applications in integrated quantum circuits \cite{kurman2026poweringquantumcomputationquantum}.

Despite these theoretical developments, the practical operation of QBs is severely constrained by dynamical instability. Realistic QBs inherently interact with surrounding thermal or Markovian environments, triggering open-system dynamics. During the charging phase, QBs typically display pronounced coherent energy backflow and non-Markovian persistent oscillations, making it difficult to lock the battery at its maximum energy capacity \cite{Zahia2025, Cavaliere2025}. Furthermore, environmental dissipation causes rapid discharge over extended timescales \cite{bhyh-53np}, significantly degrading the net extractable work, or ergotropy \cite{Hymas2026}. Simply scaling system dimensions or increasing driving strengths does not systematically resolve the trade-off between fast energy transfer and long-term storage stability.

\begin{figure}[b]
  \centering
  \includegraphics[width=0.45\textwidth]{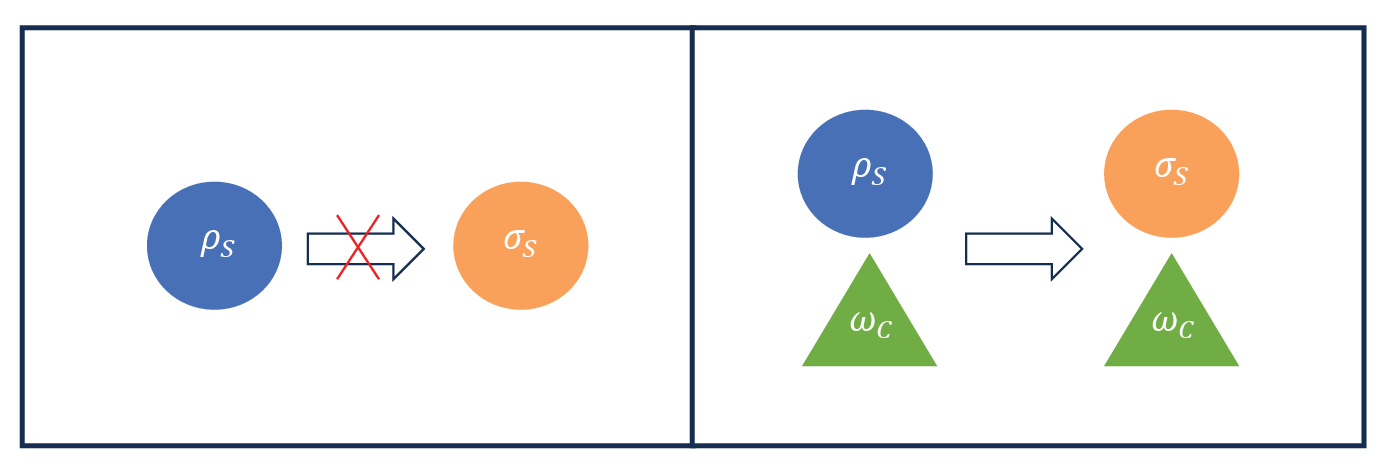}
  \caption{Conceptual schematic of Quantum Catalysis. Within the framework of quantum resource theory, a direct state transformation ${\rho}_{S} \rightarrow {\sigma}_{S}$ may be strictly forbidden by operational constraints or conservation laws (left panel). However, this targeted transformation becomes physically attainable by introducing an ancillary catalytic system in state ${\omega}_{c}$, realizing the joint evolution ${\rho}_{S} \otimes {\omega}_{C} \rightarrow {\sigma}_{S} \otimes {\omega}_{C}$ (right panel), where the catalyst remains unconsumed and globally unaltered.
  }\label{Cat act}
\end{figure}

To address these dynamic bottlenecks, recent attention has turned toward quantum catalysis-a framework originating from quantum resource theory(As illustrated in Fig.\ref{Cat act}) \cite{PhysRevLett.117.030401,PhysRevLett.83.3566}. In the context of quantum thermodynamic devices and QBs, existing studies on quantum catalysis broadly fall into three principal directions: closed-system catalytic resource protocols focusing on kinematically enabling state transformations under strict conservation bounds \cite{PhysRevA.107.042419}; correlated catalytic charging schemes that exploit transient system-catalyst entanglement to accelerate energy transfer rates \cite{PhysRevB.108.L180301, Fang2025}; and thermal catalytic cycles using auxiliary qudits to modulate heat flows in finite-time thermodynamic engines \cite{PhysRevLett.133.250201, PhysRevLett.132.260403}.

While these approaches demonstrate the versatility of catalysis, significant physical limitations persist across each domain. Closed-system protocols rely heavily on ideal, isolated dynamics, leaving their theoretical bounds vulnerable to environmental decoherence. Correlated catalytic charging schemes frequently tolerate substantial transient energy occupation within the catalyst or residual system-catalyst correlations, which obscures whether performance gains originate from pure catalytic mediation or implicit auxiliary energy injection. Meanwhile, thermal catalytic frameworks primarily target asymptotic steady-state efficiency rather than mitigating the transient coherent backflow that directly destabilizes battery charging. Consequently, a unified microscopic dynamical mechanism capable of suppressing energy backflow while preserving strict catalyst invariance in dissipative environments remains to be fully clarified.

Motivated by these open questions, this work investigates a catalytic charging architecture explicitly engineered to stabilize energy storage and mitigate backflow in open quantum batteries. In quantum resource dynamics, quantum catalysts generally fall into two distinct physical paradigms: \textit{state-invariant catalysts}, defined by strict trace-preserving state invariance $\rho_C(f) = \rho_C(0)$, and \textit{energy-conserving catalysts}, defined by neutral expectation value dynamics $\langle H_C(t) \rangle = \langle H_C(0) \rangle$ without net energy accumulation. Here, we introduce an auxiliary catalytic mode into a driver-battery network, employing the extended Lindblad master equation to trace the joint open-system dynamics. We analyze how the catalyst acts as a passive, invariant energy-transfer conduit that suppresses backflow without retaining transient excitation energy or corrupting its initial state. The ergotropy dynamics are evaluated to assess the stability of the extractable work under continuous dissipation.

The remainder of this paper is structured as follows. Section~\ref{Model} introduces the theoretical model, including the microscopic Hamiltonians for both standalone and catalytic configurations, as well as the open-system master equation formalism. Section~\ref{Results} presents numerical results on charging dynamics, with a focus on energy retention, backflow suppression, and scaling characteristics. Section~\ref{Discussions} analyzes the underlying energy-transfer mechanism and validates the realizability of the scheme on state-of-the-art physical platforms. Section~\ref{Experimental} further discusses feasible experimental implementations constrained by the parameter range of the established theoretical model. Finally, Section~\ref{conclusions} concludes the key results and elaborates their implications for robust quantum thermodynamic architectures.

\section{Theoretical Framework and Quantum Battery Models}\label{Model}

In this section, we formulate the microscopic theoretical framework governing the charging dynamics of the quantum battery (QB) system. We first construct the conventional bipartite charger-battery architecture as a benchmark and subsequently introduce the tripartite setup featuring a quantum catalyst that mediates energy transfer between the charger and the battery.

\subsection{Unassisted Bipartite Quantum Battery}

\begin{figure}[b]
  \centering
  \includegraphics[width=0.46\textwidth]{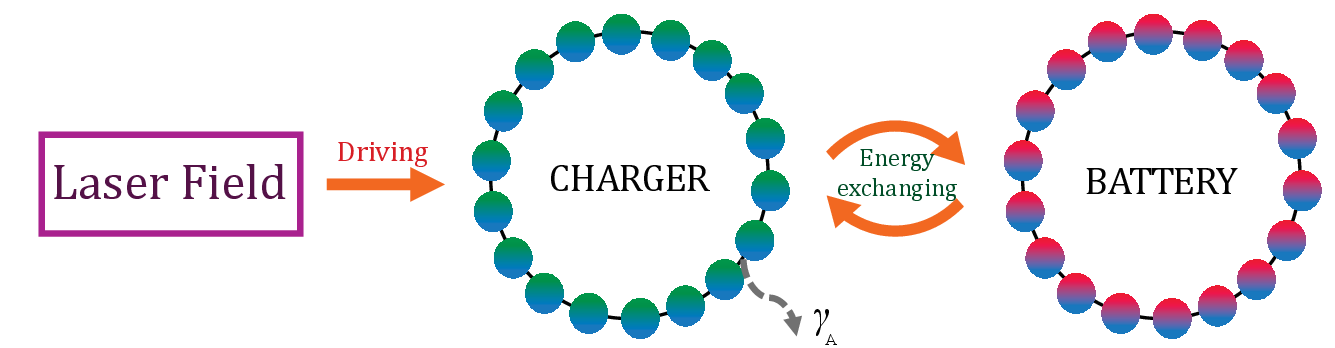}
  \caption{Schematic of the conventional bipartite quantum battery system. A classical optical field drives the charger ($A$), which subsequently transfers energy to the target battery ($B$) via direct collective coupling $J_1$. The charger is subjected to irreversible energy dissipation into a thermal environment (wavy arrow). Internal nearest-neighbor exchange interactions $J_0$ are active within both $A$ and $B$ under periodic boundary conditions.}
  \label{without cat}
\end{figure}

We consider a bipartite quantum battery system comprising a charger ($A$) and a battery ($B$), composed of $N_A$ and $N_B$ identical two-level atoms (qubits), respectively, as schematically illustrated in Fig.~\ref{without cat}.

The charger is continuously driven by a classical monochromatic optical field with frequency $\omega_f$ and amplitude $F$. Assuming the spatial dimensions of both atomic ensembles are small compared to the wavelength of the driving field-corresponding to the Dicke limit \cite{PhysRev.93.99}-the spatial variation of the field is neglected, and the collective behavior of the atoms is described by the collective spin operators $\hat{\sigma}_{\alpha}^K = \sum\limits_{i=1}^{N_K} \hat{\sigma}_{\alpha}^{(i)}$, where $K \in \{A, B\}$ and $\alpha \in \{x, y, z, +, -\}$. These collective operators obey the standard $\mathrm{SU}(2)$ commutation relations $[\hat{\sigma}_{z}^K, \hat{\sigma}_{\pm}^K] = \pm 2\hat{\sigma}_{\pm}^K$ and $[\hat{\sigma}_{+}^K, \hat{\sigma}_{-}^K] = \hat{\sigma}_{z}^K$.

Both the charger and the battery are arranged in a one-dimensional ring topology with periodic boundary conditions \cite{huang1988solid}. Setting $\hbar = 1$, the total Hamiltonian of the unassisted bipartite system in the laboratory frame is given by
\begin{equation} \label{H_total_lab_uncat}
\hat{H}_{\text{uncat}}(t) = \hat{H}_0 + \hat{H}^{\text{in}} + \hat{H}_{\text{drive}}(t) + \hat{H}_{AB},
\end{equation}
where the non-interacting Hamiltonian reads
\begin{equation} \label{H0_uncat}
\hat{H}_0 = \frac{\omega_a}{2}\hat{\sigma}_z^A + \frac{\omega_b}{2}\hat{\sigma}_z^B,
\end{equation}
with $\omega_a$ and $\omega_b$ denoting the transition frequencies of the individual two-level systems in the charger and the battery. The internal dipole-dipole exchange Hamiltonian $\hat{H}^{\text{in}} = \hat{H}_A^{\text{in}} + \hat{H}_B^{\text{in}}$ accounts for nearest-neighbor interactions within each subsystem:
\begin{equation} \label{Hin_uncat}
\hat{H}_K^{\text{in}} = J_0 \sum_{i=1}^{N_K} \left( \hat{\sigma}_+^{(i)} \hat{\sigma}_-^{(i+1)} + \hat{\sigma}_-^{(i)} \hat{\sigma}_+^{(i+1)} \right), \quad (K \in \{A, B\}),
\end{equation}
with boundary identification $\hat{\sigma}_{\pm}^{(N_K+1)} \equiv \hat{\sigma}_{\pm}^{(1)}$, where $J_0$ represents the intra-subsystem coupling strength.

The external laser driving on the charger and the resonant energy exchange between the charger and battery are described by
\begin{align}
\hat{H}_{\text{drive}}(t) &= F \left( e^{-i\omega_f t}\hat{\sigma}_+^A + e^{i\omega_f t}\hat{\sigma}_-^A \right), \label{Hdrive} \\
\hat{H}_{AB} &= J_1 \left( \hat{\sigma}_+^A \hat{\sigma}_-^B + \hat{\sigma}_-^A \hat{\sigma}_+^B \right), \label{HAB}
\end{align}

\noindent where $J_1$ is the inter-subsystem coupling strength. To eliminate explicit time dependence, we transform the system into a rotating frame defined by the unitary operator $\hat{U}_1(t) = \exp[-i\omega_f t (\hat{\sigma}_z^A + \hat{\sigma}_z^B)/2]$. Under the rotating-wave approximation (RWA) \cite{10.1093/oso/9780198501770.001.0001}, the effective time-independent Hamiltonian becomes
\begin{equation} \label{H_eff_uncat}
\hat{H}_{\text{eff}} = \frac{\Delta_a}{2}\hat{\sigma}_z^A + \frac{\Delta_b}{2}\hat{\sigma}_z^B + \hat{H}^{\text{in}} + F(\hat{\sigma}_+^A + \hat{\sigma}_-^A) + \hat{H}_{AB},
\end{equation}

\noindent where $\Delta_a = \omega_a - \omega_f$ and $\Delta_b = \omega_b - \omega_f$ represent the detunings of the charger and battery relative to the driving frequency.

To model dissipative dynamics during charging, we assume the battery functions as an ideal storage cavity with negligible losses, whereas the charger is coupled to a Markovian thermal reservoir \cite{10.1093/acprof:oso/9780199213900.001.0001}. The time evolution of the density matrix $\hat{\rho}_{AB}(t)$ is governed by the Lindblad master equation:
\begin{equation} \label{master_uncat}
\dot{\hat{\rho}}_{AB} = -i [\hat{H}_{\text{eff}}, \hat{\rho}_{AB}] + \gamma_a (N_a + 1) \mathcal{D}[\hat{\sigma}_-^A] \hat{\rho}_{AB} + \gamma_a N_a \mathcal{D}[\hat{\sigma}_+^A] \hat{\rho}_{AB},
\end{equation}

\noindent where $\mathcal{D}[\hat{\mathcal{O}}] \hat{\rho} = \hat{\mathcal{O}} \hat{\rho} \hat{\mathcal{O}}^\dag - \frac{1}{2} \{ \hat{\mathcal{O}}^\dag \hat{\mathcal{O}}, \hat{\rho} \}$ denotes the standard Lindbladian dissipator. The parameter $\gamma_a$ characterizes the spontaneous relaxation rate of the charger, and $N_a = [\exp(\omega_a / k_B T) - 1]^{-1}$ is the mean thermal photon number at temperature $T$.

\subsection{Catalytic Tripartite Quantum Battery}

\begin{figure}[htb]
  \centering
  \includegraphics[width=0.46\textwidth]{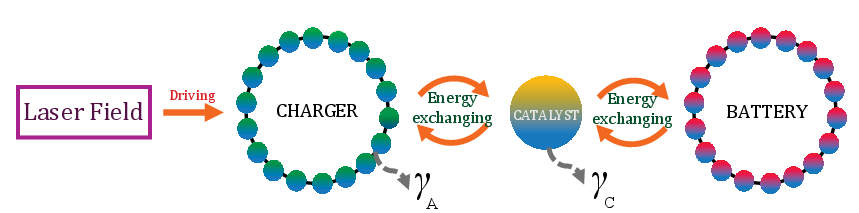}
  \caption{Schematic of the catalyst-mediated quantum battery system. An auxiliary catalytic system ($C$) is coupled symmetrically to the laser-driven charger ($A$) and target battery ($B$) with coupling strength $J_1$, while direct $A$-$B$ coupling is absent. Both $A$ and $C$ experience energy dissipation into local thermal environments with decay rates $\gamma_a$ and $\gamma_c$.}
  \label{with cat}
\end{figure}

We extend the bipartite model by interposing an auxiliary quantum system-the catalyst ($C$)-between the charger and the battery, as depicted in Fig.~\ref{with cat}. Direct interaction between the charger and battery is assumed to be suppressed ($J_{AB} = 0$), rendering energy transfer strictly catalyst-mediated. 

The catalyst is initially modeled as a two-level system with eigenfrequency $\omega_c$ and free Hamiltonian
\begin{equation} \label{HC_free}
\hat{H}_C = \frac{\omega_c}{2}\hat{\sigma}_z^C,
\end{equation}

\noindent where $\hat{\sigma}_z^C$ is the Pauli-$Z$ operator acting on the catalyst Hilbert space. Assuming a symmetric interaction scheme where the catalyst couples to both the charger and battery with identical strength $J_1$, the interaction Hamiltonian in the laboratory frame is
\begin{equation} \label{Hint_cat_lab}
\hat{H}_{\text{int}}^{\text{cat}}(t) = \hat{H}_{\text{drive}}(t) + J_1 \left( \hat{\sigma}_+^A\hat{\sigma}_-^C + \hat{\sigma}_-^A\hat{\sigma}_+^C + \hat{\sigma}_+^B\hat{\sigma}_-^C + \hat{\sigma}_-^B\hat{\sigma}_+^C \right).
\end{equation}

Transforming to the rotating frame via $\hat{U}_2(t) $=$ \exp[-i\omega_f t (\hat{\sigma}_z^A + \hat{\sigma}_z^B + \hat{\sigma}_z^C)/2]$ and applying the RWA yields the effective time-independent Hamiltonian:
\begin{equation} \label{H_eff_cat}
\hat{H}_{\text{eff}}^{\text{cat}} = \frac{\Delta_a}{2}\hat{\sigma}_z^A + \frac{\Delta_b}{2}\hat{\sigma}_z^B + \frac{\Delta_c}{2}\hat{\sigma}_z^C + \hat{H}^{\text{in}} + F(\hat{\sigma}_+^A + \hat{\sigma}_-^A) + \hat{H}_{\text{med}},
\end{equation}

\noindent where $\Delta_c = \omega_c - \omega_f$ represents the catalyst detuning, and $\hat{H}_{\text{med}} $=$ J_1 (\hat{\sigma}_+^A\hat{\sigma}_-^C + \hat{\sigma}_+^B\hat{\sigma}_-^C + \text{h.c.})$ describes the catalyst-mediated excitation exchange (Detailed derivations are given in Appendix ~\ref{APPENDIX}).

Allowing for open-system decoherence on both the charger and the catalyst, the full density matrix $\hat{\rho}_{ABC}(t)$ evolves according to the extended Lindblad master equation:
\begin{equation} \label{master_cat}
\begin{aligned}
\dot{\hat{\rho}}_{ABC} = -i [\hat{H}_{\text{eff}}^{\text{cat}}, \hat{\rho}_{ABC}] 
&+ \sum_{k \in \{A, C\}} \gamma_k (N_k + 1) \mathcal{D}[\hat{\sigma}_-^k] \hat{\rho}_{ABC} \\
&+ \sum_{k \in \{A, C\}} \gamma_k N_k \mathcal{D}[\hat{\sigma}_+^k] \hat{\rho}_{ABC},
\end{aligned}
\end{equation}

\noindent where $\gamma_c$ denotes the spontaneous decay rate of the catalyst, and $N_c = [\exp(\omega_c / k_B T) - 1]^{-1}$ is the thermal occupation number of its local environment.

This formalism straightforwardly maps to continuous-variable catalysts (e.g., single-mode cavity resonators or nanomechanical modes). For a bosonic catalyst, the spin operators are mapped via $\hat{\sigma}_-^C \to \hat{a}$, $\hat{\sigma}_+^C \to \hat{a}^\dag$, and $\hat{\sigma}_z^C \to 2\hat{a}^\dag\hat{a}$, where $[\hat{a}, \hat{a}^\dag] = 1$, preserving the structural algebraic form of Eq.~\eqref{master_cat}.

\subsection{Thermodynamic Metrics: Ergotropy and Energy Currents}

To evaluate the usable energy stored in the quantum battery, we compute the ergotropy $W_B(t)$ \cite{Choquehuanca_2025,PhysRevA.104.L030402}, defined as the maximum work extractable via cyclic unitary operations:
\begin{equation} \label{ergotropy_def}
W_B(t) = \operatorname{Tr} \left[ \hat{H}_B^{\text{tot}} \hat{\rho}_B(t) \right] - \min_{\hat{U}_B} \operatorname{Tr} \left[ \hat{H}_B^{\text{tot}} \hat{U}_B \hat{\rho}_B(t) \hat{U}_B^\dag \right],
\end{equation}

\noindent where $\hat{H}_B^{\text{tot}} $=$ \frac{\omega_b}{2}\hat{\sigma}_z^B + \hat{H}_B^{\text{in}}$ represents the total local battery Hamiltonian, $\hat{\rho}_B(t) $=$ \operatorname{Tr}_{AC}[\hat{\rho}_{ABC}(t)]$  (or $\operatorname{Tr}_{A}[\hat{\rho}_{AB}(t)]$ in the absence of the catalyst) is the reduced density matrix of the battery, and $\hat{U}_B$ denotes an arbitrary unitary transformation acting solely on the battery Hilbert space. 

To bridge the dynamical quantities in the rotating frame with the thermodynamic observables in the laboratory frame, we employ the unitary transformation $\hat{U}(t)$. The laboratory-frame density matrix $\hat{\rho}_{\text{lab}}(t)$ and total Hamiltonian $\hat{H}_{\text{lab}}(t)$ are related to their rotating-frame counterparts, $\hat{\rho}_{\text{rot}}(t)$ and $\hat{H}_{\text{eff}}^{\text{(cat,uncat)}}$, via:
\begin{subequations} \label{frame_transformations}
\begin{align}
\hat{\rho}_{\text{lab}}(t) &= \hat{U}^\dag(t) \hat{\rho}_{\text{rot}}(t) \hat{U}(t), \label{rho_transform} \\
\hat{H}_{\text{lab}}(t) &= \hat{U}^\dag(t) \hat{H}_{\text{eff}} \hat{U}(t) + i \left( \frac{d\hat{U}^\dag(t)}{dt} \right) \hat{U}(t). \label{H_transform}
\end{align}
\end{subequations}

Because the battery local Hamiltonian commutes with the generator of the frame transformation, i.e., $[\hat{H}_B^{\text{tot}}, \hat{U}(t)] = 0$, the local spectrum and passive states are invariant under $\hat{U}(t)$. Consequently, the ergotropy $W_B(t)$ defined in Eq.~\eqref{ergotropy_def} yields identical numerical values whether evaluated using $\hat{\rho}_{\text{lab}}(t)$ or $\hat{\rho}_{\text{rot}}(t)$.

Similarly, for the mediating catalyst, its transient energy occupation is monitored via
\begin{equation} \label{catalyst_energy}
E_C(t) = \operatorname{Tr} \left[ \hat{H}_C \hat{\rho}_C(t) \right],
\end{equation}

\noindent where $\hat{\rho}_C(t) = \operatorname{Tr}_{AB}[\hat{\rho}_{ABC}(t)]$, which is likewise frame-invariant due to $[\hat{H}_C, \hat{U}(t)] = 0$.

Furthermore, to characterize the non-equilibrium energy exchange during charging within the standard Alicki thermodynamic framework \cite{Alicki1979,PhysRevA.74.063823,Oh2020}, the rate of change of the total system energy $E(t) = \operatorname{Tr}[\hat{H}_{\text{lab}}(t) \hat{\rho}_{\text{lab}}(t)]$ in the laboratory frame is partitioned into the external driving power $P(t)$ and the heat current $J(t)$:
\begin{equation} \label{first_law}
\frac{d}{dt} E(t) = P(t) + J(t).
\end{equation}
By substituting Eqs.~\eqref{rho_transform}--\eqref{H_transform} into the standard definitions, the instantaneous driving power injected by the external classical laser field is expressed as
\begin{equation} \label{power_def}
P(t) \equiv \operatorname{Tr} \left[ \hat{\rho}_{\text{lab}}(t) \frac{\partial \hat{H}_{\text{lab}}(t)}{\partial t} \right] 
\end{equation}

\noindent while the irreversible heat current flowing into the thermal reservoirs, driven by the non-unitary open-system dynamics, is given by
\begin{align} \label{heat_current_def}
J(t) \equiv \operatorname{Tr} \left[  \frac{d\hat{\rho}_{\text{lab}}(t)}{dt} \hat{H}_{\text{lab}}(t) \right] 
\end{align}

Equations~\eqref{power_def} and \eqref{heat_current_def} establish an exact physical equivalence between the rotating-frame evolution and laboratory-frame thermodynamics. Tracking $W_B(t)$, $E_C(t)$, $P(t)$, and $J(t)$ thus provides a comprehensive and mathematically rigorous description of energy conversion efficiency, work extractability, and catalytic stability.

\section{Numerical Results and Physical Mechanisms}\label{Results}

\begingroup\squeezetable
\begin{table}[h]
    \centering                       
    \caption{Simulation parameter values in this work.}       
    \begin{tabular}{l l}
        \toprule
        Parameter & Value \\
        \midrule
        The eigenfrequencies of the charger $\omega_a$ & $\omega$  \\
        The eigenfrequencies of the battery $\omega_b$ & $\omega$  \\
        The eigenfrequencies of the catalyst $\omega_c$ & $0.1\omega$  \\
        The frequency of the driving optical field $\omega_f$ & $\omega$  \\
        The amplitude of the driving optical field $F$ & $0.1\omega$ \\
        The coupling strength between atoms within the device $J_0$      & $0.1\omega$ \\
        The coupling strength between different devices  $J_1$ & $0.3\omega$ \\
        The dissipation rate of the charger $\gamma_a$ & $0.1\omega$ \\
        The dissipation rate of the catalyst $\gamma_c$ & $0.1\omega$ \\
        The environmental temperature $T$               & 0\\
        \bottomrule
    \end{tabular}
    \label{table}
\end{table}\endgroup

To systematically evaluate the performance and underlying mechanism of the proposed catalytic quantum battery (QB) architecture, we numerically integrate the open-system dynamics formulated in Sec.~\ref{Model}. Our analysis focuses on three core quantities: the extractable work defined by the battery ergotropy $W_B(t)$ [Eq.~\eqref{ergotropy_def}], the transient catalytic energy occupation $E_C(t)$ [Eq.~\eqref{catalyst_energy}], and the irreversible heat dissipation current $J(t)$ [Eq.~\eqref{heat_current_def}]. Specifically, tracking $E_C(t)$ serves as a crucial benchmark to confirm that the auxiliary system $C$ functions strictly as a catalyst--mediating energy transfer without net energy accumulation or persistent population trapping. Furthermore, inspecting the instantaneous heat current $J(t)$ provides microscopic insight into how catalytic coupling suppresses environmental dissipation and coherent backflow, thereby stabilizing energy storage.

Unless stated otherwise, the physical parameters governing the charger, catalyst, and battery subsystems are selected within realistic experimental regimes and summarized in Table~\ref{table}.

\begin{figure*}
\centering   
{
\begin{minipage}[b]{0.22\linewidth} 
	\centering
	\includegraphics[scale=0.45]{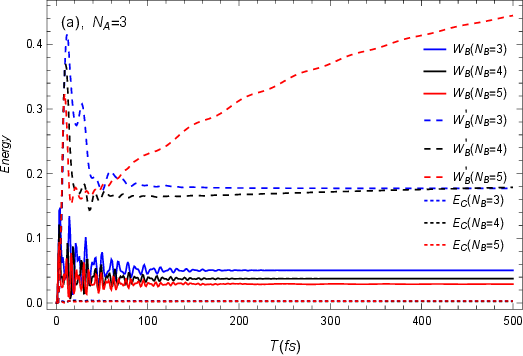}
\end{minipage}\label{group one a}}
{
\begin{minipage}[b]{0.22\linewidth}
	\centering
	\includegraphics[scale=0.45]{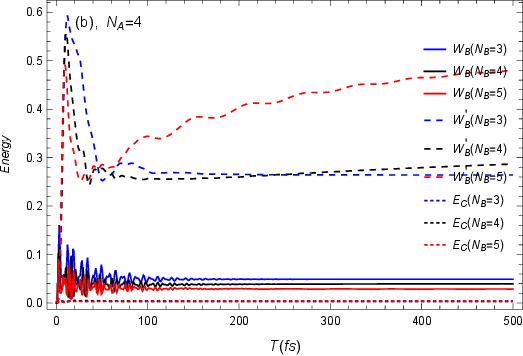}
\end{minipage}\label{group one b}}
{
\begin{minipage}[b]{0.22\linewidth}
    \centering
    \includegraphics[scale=0.45]{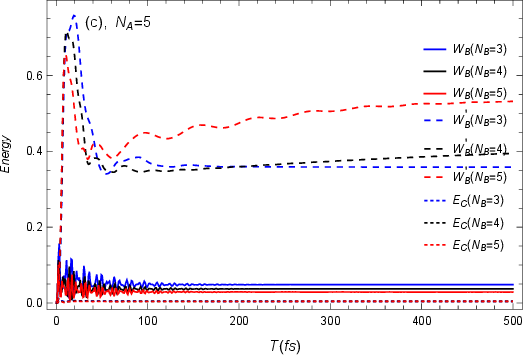}
	\end{minipage}\label{group one c}}
{
\begin{minipage}[b]{0.22\linewidth}
\centering
	\includegraphics[scale=0.45]{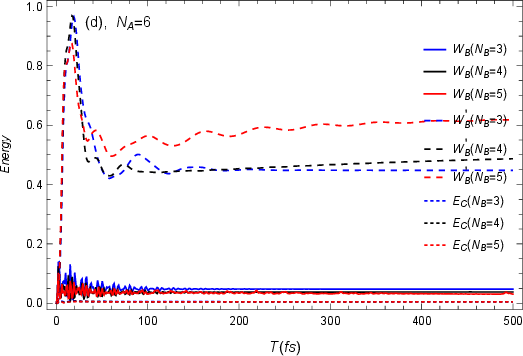}
	\end{minipage}\label{group one d}}
\caption{Time evolution of the battery ergotropy and catalyst energy. Dashed and Solid curves correspond to the battery ergotropy with ($W'_B$) and without ($W_B$) the catalyst, respectively; dotted curves denote the catalyst energy ($E_C$). Panels (a)--(d) represent charger sizes $N_A = 3, 4, 5$, and $6$. Within each panel, blue, black, and red lines indicate battery sizes $N_B = 3, 4$, and $5$, respectively. Other parameters are specified in Table~\ref{table}.}\label{energy}
\end{figure*}

To evaluate the operational performance and energy-storage mechanisms of the catalytic architecture, we systematically analyze the temporal profile of the extractable work alongside the energy occupation of the auxiliary mode. Fig.~\ref{energy} contrasts the time evolution of the battery ergotropy with ($W'_B$, dashed curves) and without ($W_B$, solid curves) the catalyst against the transient catalyst energy ($E_C$, dotted curves) across various charger sizes ($N_A = 3, 4, 5, 6$) and battery capacities ($N_B = 3, 4, 5$).

A defining characteristic of quantum catalysis is the dynamical energy invariance of the auxiliary degree of freedom. As delineated in all panels of Fig.~\ref{energy}, the dotted trajectories corresponding to different battery sizes ($N_B = 3, 4, 5$) collapse identically onto a flat, near-zero energy baseline throughout the charging sequence, 
remains essentially unchanged over the investigated range of  $N_A$ and $N_B$. This strict invariance indicates that the auxiliary system acts purely as an energy-neutral conduit, facilitating energy exchange between the driver and the battery without persistent excitation trapping or net energy consumption\cite{Chaki2025}.

The physical advantage of this catalytic mediation becomes manifest upon comparing the battery ergotropy profiles. In the unassisted bipartite model ($W_B$, solid curves), the ergotropy undergoes high-frequency coherent oscillations during the initial charging stage, reflecting reversible excitation exchange that destabilizes energy retention and compromises net work extraction. Upon interposing the catalyst ($W'_B$, dashed curves), these transient oscillations are substantially quenched, enabling a rapid and smooth convergence toward a stationary state. Furthermore, the asymptotic ergotropy plateaus achieved in the presence of the catalyst consistently exceed the peak values observed in the unassisted counterpart across all subsystem configurations. Rather than serving as an additional energy reservoir, the auxiliary system modifies the microscopic excitation-transfer dynamics while preserving its own energy occupation. This catalytic modulation suppresses transient oscillatory behavior and facilitates the formation of a stable charging regime with enhanced steady-state ergotropy of the many-body quantum battery.

\begin{figure*}[t]
\hspace{-3cm}
{\begin{minipage}[b]{0.2\linewidth} 
	\centering
	\includegraphics[scale=0.5]{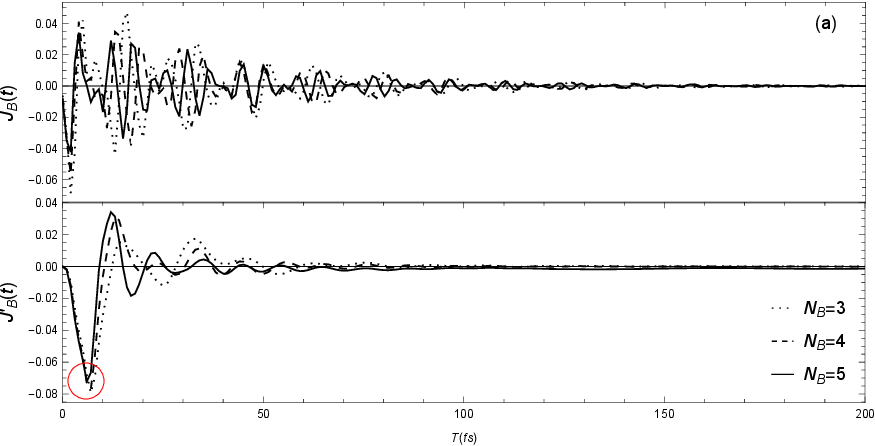}
\end{minipage}\label{JB a}\hspace{4cm}
\begin{minipage}[b]{0.2\linewidth}
	\centering
	\includegraphics[scale=0.5]{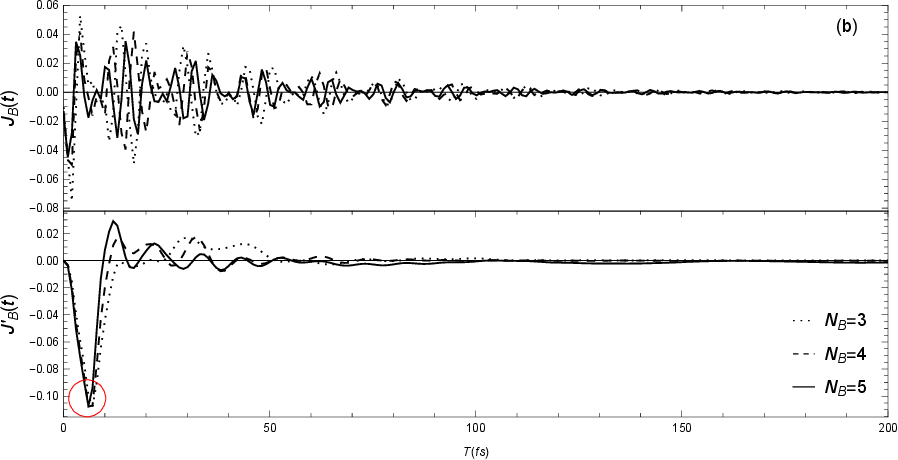}
\end{minipage}\label{JB b}}\vspace{0.25cm}

\hspace{-3cm}	
{\begin{minipage}[b]{0.2\linewidth}
	\centering
	\includegraphics[scale=0.5]{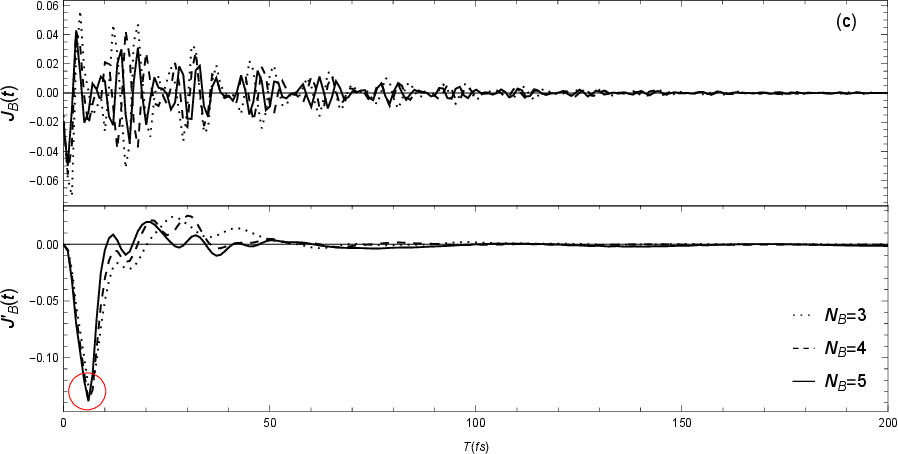}
\end{minipage}\label{JB c}\hspace{4cm}
\begin{minipage}[b]{0.2\linewidth}
	\centering
	\includegraphics[scale=0.5]{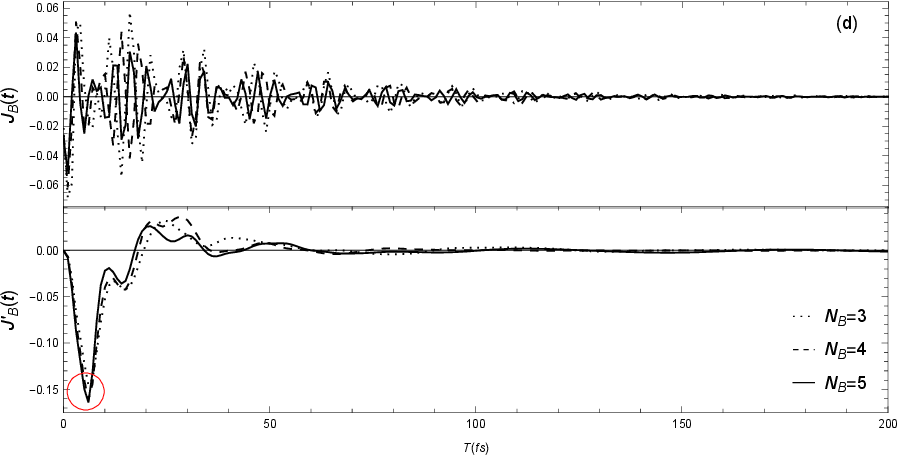}
\end{minipage}}\label{JB d}
\caption{Time evolution of the quantum battery heat current without ($J_B(t)$) and with ($J'_B(t)$) the catalyst. Panels (a)--(d) correspond to charger sizes $N_A = 3, 4, 5$, and $6$. Dotted, dashed, and solid curves denote battery sizes $N_B = 3, 4$, and $5$, respectively. Parameters are identical to those in Fig.~\ref{energy}.}\label{heatcurrent}
\end{figure*}

To elucidate how the auxiliary catalytic system modifies the energy exchange dynamics of the many-body quantum battery, we examine the transient behavior of the heat current and its direct impact on the storage and redistribution of ergotropy. We first specify the sign convention for the irreversible heat dissipation current $J(t)$ defined in Eq.~\eqref{heat_current_def}. In accordance with standard quantum thermodynamic conventions, energy transferred into the system is defined as negative ($J(t) < 0$), whereas outward energy flow is positive ($J(t) \ge 0$). Consequently, for the battery heat current, $J_B(t) < 0$ signifies net energy charging from the charger to the battery, while $J_B(t) \ge 0$ denotes energy backflow toward the charger or the environment. For clarity, the heat currents in the presence and absence of the catalyst are denoted by $J'_B(t)$ and $J_B(t)$, respectively.

Fig.~\ref{heatcurrent} illustrates the temporal evolution of $J_B(t)$ and $J'_B(t)$ over a $200\text{~fs}$ interval for various charger sizes ($N_A = 3, 4, 5, 6$) and battery sizes ($N_B = 3, 4, 5$). In the unassisted case, $J_B(t)$ exhibits strong transient oscillations that progressively decay toward zero, reflecting periodic coherent energy exchange between the battery and charger before reaching long-time thermalization. In contrast, the catalytic current $J'_B(t)$ develops a prominent negative minimum at early times, followed by significantly attenuated oscillations and subsequent relaxation toward a stationary plateau. This enhanced negative transient region ( denoted by the red circles in Panels (a)-(d) of Fig.~\ref{heatcurrent}) indicates that the catalyst accelerates the early-stage energy injection while suppressing the subsequent coherent energy backflow.

This catalytic modulation of the heat-current dynamics directly governs the capacity and stability of energy storage within the battery. As shown in Fig.~\ref{energy}, the early-time surge in energy intake correlates with higher initial peaks in the total stored energy. Importantly, the suppression of large-amplitude backflow oscillations prevents periodic depletion of stored energy, thereby shifting the ergotropy dynamics toward higher baseline values. As a consequence, the asymptotic ergotropy plateaus achieved with the catalyst consistently surpass the maximum transient peaks observed in the unassisted counterpart across all evaluated subsystem geometries. These findings demonstrate that the auxiliary catalyst enhances the performance of many-body quantum batteries by rerouting the transient dissipation pathways, ultimately stabilizing a high-ergotropy charging state.

\section{Discussions: Microscopic Mechanism of Catalytic Control}\label{Discussions}

To elucidate the theoretical origin of the backflow suppression in $J_B(t)$ and the stabilization of $W_B(\infty)$, we formulate a microscopic analysis using Liouvillian spectral decomposition and effective Hamiltonian engineering in the dispersive regime.

\subsection{Energy-Neutral Condition of the Catalytic Mode}

A key requirement for quantum catalysis is that the auxiliary system facilitates energy transport without maintaining steady-state excitation. In our setup, the catalytic transition frequency is detuned from the driving frequency $\omega_f $=$ \omega_a $=$ \omega_b $=$ \omega$ by $\Delta_c $=$ \omega_c - \omega_f $=$ -0.9\omega$. In the rotating frame generated by $\hat{U}(t) $=$ \exp[-i\omega_f t (\hat{\sigma}_z^A + \hat{\sigma}_z^B + \hat{\sigma}_z^C)/2]$, the total Hamiltonian reads
\begin{equation} \label{Hf_with_cat_disc}
\hat{H}' = F \hat{\sigma}_x^A + \frac{\Delta_c}{2} \hat{\sigma}_z^C + \hat{H}^{\text{in}} + J_1 \left( \hat{\sigma}_-^A \hat{\sigma}_+^C + \hat{\sigma}_-^B \hat{\sigma}_+^C + \text{h.c.} \right).
\end{equation}
Under the dispersive hierarchy $|\Delta_c| \gg \{J_1, F, \gamma_c\}$, real excitations of mode $C$ are off-resonantly suppressed.

Applying steady-state perturbation theory to the master equation, the stationary excited-state population $P_{ee}^C = \langle \hat{\sigma}_+^C \hat{\sigma}_-^C \rangle_{\text{ss}}$ is bounded by
\begin{equation} \label{the_population_of_the_C_excited_state}
P_{ee}^C \approx \frac{J_1^2 |\langle \hat{\sigma}_-^A \rangle + \langle \hat{\sigma}_-^B \rangle|^2}{\Delta_c^2 + (\gamma_c / 2)^2} \le \frac{J_1^2}{\Delta_c^2 + (\gamma_c / 2)^2}.
\end{equation}
For $J_1 = 0.3\omega$ and $\Delta_c = -0.9\omega$, Eq.~\eqref{the_population_of_the_C_excited_state} yields $P_{ee}^C \sim O(10^{-2})$. This upper bound confirms that population accumulation within $C$ remains negligible, restricting the auxiliary mode to virtual state transitions that mediate interaction without long-time energy storage.

\subsection{Effective Complex Coupling and Dissipative Backflow Suppression}

In the unassisted bipartite setting ($J_{AB} = J_1$), system dynamics are governed by the superoperator $\mathcal{L}_{\text{uncat}} \hat{\rho}_{AB} = -i[\hat{H}_{\text{uncat}}, \hat{\rho}_{AB}] + \gamma_a \mathcal{D}[\hat{\sigma}_-^A]\hat{\rho}_{AB}$. The strong exchange coupling $J_1 = 0.3\omega$ relative to the dissipation rate $\gamma_a = 0.1\omega$ places the $A\text{--}B$ channel in the underdamped regime ($J_1 > \gamma_a / 2$). The Liouvillian spectrum contains imaginary eigenvalues $\text{Im}(\lambda_k) \propto J_1$ with decay rates $\text{Re}(\lambda_k) \approx -\gamma_a / 2$, resulting in persistent coherent backflow from $B$ to $A$ and periodic depletion of the battery population inversion.

In the catalytic architecture ($J_{AB} = 0$), adiabatic elimination of mode $C$ up to second order in $J_1 / |\Delta_c|$ yields an effective bipartite Liouvillian acting on $A \otimes B$. The mediated interaction induces an effective complex coupling:
\begin{equation} \label{J_eff}
J_{\text{eff}} = \frac{J_1^2}{- \Delta_c + i \gamma_c / 2} = \operatorname{Re}(J_{\text{eff}}) + i \operatorname{Im}(J_{\text{eff}}),
\end{equation}
where
\begin{align}
\operatorname{Re}(J_{\text{eff}}) &= -\frac{J_1^2 \Delta_c}{\Delta_c^2 + \gamma_c^2/4}, \label{Re_Jeff} \\
\operatorname{Im}(J_{\text{eff}}) &= -\frac{J_1^2 \gamma_c / 2}{\Delta_c^2 + \gamma_c^2/4}. \label{Im_Jeff}
\end{align}
Additionally, second-order terms generate local AC Stark shifts $\operatorname{Re}(J_{\text{eff}}) (\hat{\sigma}_+^A \hat{\sigma}_-^A + \hat{\sigma}_+^B \hat{\sigma}_-^B)$ that shift the effective subsystem frequencies.

The impact of $J_{\text{eff}}$ on the heat current $J'_B(t)$ is established through its explicit dependence on inter-system correlation functions:
\begin{align} \label{JB_correlation}
J'_B(t) &= -i J_1 \omega_b \operatorname{Tr} \left[ \left( \hat{\sigma}_+^B \hat{\sigma}_-^C - \hat{\sigma}_-^B \hat{\sigma}_+^C \right) \hat{\rho}_{ABC}(t) \right] \nonumber\\
&= 2 J_1 \omega_b \operatorname{Im}\langle \hat{\sigma}_+^B \hat{\sigma}_-^C \rangle(t).
\end{align}
Equations \eqref{Re_Jeff}--\eqref{JB_correlation} establish two distinct physical mechanisms:
\begin{enumerate}
    \item \textbf{Shift Toward Critically Damped Regime:} The real effective exchange rate is renormalized to $\vert{}Re(J_{\text{eff}})\vert{}\approx0.1\omega$. In the unassisted bipartite setup, the system operates deep within the underdamped regime, as the direct coupling strength far exceeds the dissipation rate ($2J_1 = 0.6\omega \gg \gamma_a = 0.1\omega$). Upon introducing the catalyst, the effective interaction strength is reduced such that $2\vert{}Re(J_{\text{eff}})\vert{} \approx 0.2\omega$, causing the joint dynamics to move much closer to the critically damped threshold ($2\vert{}Re(J_{\text{eff}})\vert{} \to \gamma_a$). This shift effectively quenches high-frequency transient Rabi oscillations without inducing sluggish overdamped dynamics.       
     \item \textbf{Non-Hermitian Coherence Damping:} The imaginary term $\operatorname{Im}(J_{\text{eff}})$ introduces a non-Hermitian damping mechanism targeting transition coherences $\langle \hat{\sigma}_+^A \hat{\sigma}_-^B \rangle$ and $\langle \hat{\sigma}_+^B \hat{\sigma}_-^C \rangle$. Because this damping targets off-diagonal phase coherence rather than diagonal populations $\langle \hat{\sigma}_z^B \rangle$, it quenches backflow oscillations in $J'_B(t)$ without draining net population from battery $B$.
\end{enumerate}

\subsection{Thermodynamic Link to Steady-State Ergotropy}

The ergotropy $W_B(t)$ quantifies the maximum work extractable from the reduced state $\hat{\rho}_B(t)$ relative to its passive state $\hat{\sigma}_B^\pi$:
\begin{equation} \label{ergotropy_def}
W_B(t) = \operatorname{Tr}\left( \hat{H}_B \hat{\rho}_B(t) \right) - \operatorname{Tr}\left( \hat{H}_B \hat{\sigma}_B^\pi \right),
\end{equation}
where $\hat{\sigma}_B^\pi$ is formed by ordering state populations monotonically decreasing with energy.

Since passive state transformations cannot extract energy from diagonal population inversions, maximizing $W_B(t)$ requires maintaining population inversion $\langle \hat{\sigma}_z^B \rangle$ while suppressing mixedness $1 - \operatorname{Tr}(\hat{\rho}_B^2)$. In the unassisted configuration, persistent backflow oscillations periodically deplete $\langle \hat{\sigma}_z^B \rangle(t)$, inducing transient collapses in $W_B(t)$.

Under catalytic mediation, the suppression of backflow oscillations via $\operatorname{Im}(J_{\text{eff}})$ together with overdamped dynamics ensures that energy transferred during the initial negative current phase ($J'_B(t) < 0$) is trapped within $B$. As $t \to \infty$, the battery relaxes toward a stationary state $\hat{\rho}_B(\infty)$ defined by a locked diagonal inversion $\langle \hat{\sigma}_z^B \rangle_{\text{ss}}$. Consequently, the auxiliary catalytic mode converts energy that would otherwise be lost during backflow cycles into stably stored, extractable work, elevating the asymptotic ergotropy $W_B'(\infty)$ above the transient peaks of the unassisted architecture.

\section{Experimental Feasibility}\label{Experimental}

The catalytic quantum battery architecture proposed in this work can be realized on current quantum hardware platforms, particularly within superconducting quantum circuits and cavity QED systems.

\subsection{Physical Platform and Parameter Mapping}

Superconducting quantum circuits provide an ideal testbed for implementing the symmetric mediated coupling $J_1 (\hat{\sigma}_-^A \hat{\sigma}_+^C + \hat{\sigma}_-^B \hat{\sigma}_+^C + \text{h.c.})$ with tunable parameters \cite{PhysRevB.71.144511,1wb4-df6c,PhysRevA.107.023725}. The charger $A$ and battery $B$ can be instantiated by arrayed Transmon qubits, while the catalytic mode $C$ can be realized using either an auxiliary flux-tunable Transmon qubit \cite{PhysRevApplied.15.054001,PhysRevA.76.042319} or a high-$Q$ 3D microwave cavity \cite{PhysRevLett.125.140801}. The requisite condition $J_{AB} = 0$ is naturally satisfied or enforced via capacitive/inductive decoupling circuits or off-resonant frequency ordering.

Considering a standard fundamental frequency $\omega / 2\pi $=$ 5.0\text{~GHz}$ for Transmon qubits, the normalized parameters utilized in Secs.~\ref{Results} and \ref{Discussions} correspond to the following physical values:
\begin{itemize}
    \item Inter-system coupling strength: $J_1 / 2\pi $=$ 0.3 \omega / 2\pi = 1.5\text{~GHz}$ (achievable in the strong/ultrastrong coupling regime via gapped flux-line couplers \cite{RevModPhys.91.025005});
    \item Catalyst detuning: $\Delta_c / 2\pi = -0.9 \omega / 2\pi $=$ -4.5\text{~GHz}$ (corresponding to a catalyst frequency $\omega_c / 2\pi = 0.5\text{~GHz}$, within the wide flux-tuning bandwidth of Transmons);
    \item Charger dissipation rate: $\gamma_a / 2\pi = 0.1 \omega / 2\pi = 500\text{~MHz}$ (controlled via Purcell filters or adjustable radiative couplers \cite{PhysRevLett.101.080502}).
\end{itemize}
Alternatively, using moderate coupling strengths with a baseline frequency $\omega / 2\pi $=$ 2.0\text{~GHz}$, $J_1 / 2\pi $=$ 600\text{~MHz}$ and $\gamma_a / 2\pi $=$ 200\text{~MHz}$ fall comfortably within demonstrated experimental boundaries.

\subsection{State Preparation and Measurement Protocol}

The operational protocol relies on two experimental steps:
\begin{enumerate}
    \item \textbf{Initialization and Driving:} The qubits are initialized to their ground states via passive thermal relaxation or active reset protocols. The coherent drive $F \hat{\sigma}_x^A$ is implemented using resonant microwave pulses applied to the charger manifold.
    \item \textbf{Ergotropy Extraction:} Evaluating the battery ergotropy $W_B(t)$ requires reconstructing the reduced density matrix $\hat{\rho}_B(t)$ of the $N_B$-qubit battery. For moderate battery sizes ($N_B \le 6$), full quantum state tomography (QST) using joint readout lines can directly determine $\hat{\rho}_B(t)$ \cite{PhysRevLett.102.200402}. For larger scaling ($N_B > 6$), partial tomography focused on low-order correlation functions $\langle \hat{\sigma}_z^i \rangle$ and $\langle \hat{\sigma}_+^i \hat{\sigma}_-^j \rangle$ provides tight lower bounds on the extractable work without exponential measurement overhead \cite{g45c-ssfx}.
\end{enumerate}

These platform-specific implementations confirm that the catalyst-assisted charging protocol is accessible within state-of-the-art solid-state quantum architectures.
\section{Conclusions}\label{conclusions}
In summary, we have established a quantum-catalytic architecture designed to eliminate coherent energy backflow and stabilize charging dynamics in open many-body quantum batteries. Through a comprehensive open-system Liouvillian analysis, we demonstrate that unassisted battery configurations inherently suffer from severe non-Markovian energy backflow and persistent oscillatory instability, which severely degrade long-term energy storage. 

By introducing an auxiliary catalytic mode weakly coupled between the charger and the battery, the coherent energy transfer pathways are modified. The catalyst acts as a non-Hermitian coherence-damping channel that quenches high-amplitude transient backflow without net energy accumulation or persistent excitation trapping, strictly maintaining its dynamical energy invariance $\langle H_C(t)\rangle \approx \langle H_C(0)\rangle$. Consequently, this catalytic modulation suppresses memory effects and enables monotonic energy accumulation, resulting in a significantly elevated steady-state ergotropy $W_B(\infty)$ that scales favorably with the battery size $N_B$. Given its straightforward compatibility with state-of-the-art superconducting circuit platforms, this work highlights quantum catalysis as a robust and scalable mechanism for engineering persistent, high-performance quantum energy storage devices.
\section*{Author contributions}
S. C. Zhao conceived the idea. L. Luo performed the numerical computations and wrote the draft, and S. C. Zhao did the analysis and revised the paper.

\section{Acknowledgment}

This work is supported by the National Natural Science Foundation of China ( Grant Nos. 62065009 and 61565008 ),
Yunnan Fundamental Research Projects, China ( Grant No. 2016FB009 ) and the Foundation for Personnel training projects of Yunnan Province, China ( Grant No. KKSY201207068 ).

\section*{Data Availability Statement}

This manuscript has associated data in a data repository.[Authors' comment: All data included in this manuscript are available upon reasonable request by contacting with the corresponding author.]

 \section*{Conflict of Interest}

The authors declare that they have no conflict of interest. This article does not contain any studies with human participants or animals performed by any of the authors. Informed consent was obtained from all individual participants included in the study.
\setcounter{equation}{0}
\renewcommand{\theequation}{A\arabic{equation}}
\appendix
\section{APPENDIX}\label{APPENDIX}

\subsection{Rotating Frame Transformation of the Hamiltonian}
For the Hamiltonians of both the uncatalyzed and catalyzed cases, the rotating frame transformation can be uniformly applied, with the transformation matrix chosen as
\begin{equation}\label{Uf}
\hat{U}(t)=\exp\left[-i \omega_f t \left(\hat{\sigma}_z^A+\hat{\sigma}_z^B+\hat{\sigma}_z^C\right)/2\right].
\end{equation}
For an arbitrary time-dependent Hamiltonian $\hat{H}(t)$, it can be transformed into the rotating frame at the angular frequency $\omega_f$ via the following expression \cite{PhysRevB.99.195433}
\begin{equation}\label{rotating frame transformation}
\tilde{H}(t) = \hat{U}(t)\hat{H}(t)\hat{U}^{\dag}(t) - i \hat{U}(t)\frac{\partial }{\partial t}\hat{U}^{\dag}(t).
\end{equation}
By using the following expressions:
\begin{equation}\label{pauli operater equation 1}
\exp\left(-\frac{i\omega_f t}{2}\hat{\sigma}_z\right) (\hat{\sigma}_+) \exp\left(\frac{i\omega_f t}{2}\hat{\sigma}_z\right) =\exp\left( -i\omega_f t \right)\hat{\sigma}_+,
\end{equation}
\begin{equation}\label{pauli operater equation 2}
\exp\left(-\frac{i\omega_f t}{2}\hat{\sigma}_z\right) (\hat{\sigma}_-) \exp\left(\frac{i\omega_f t}{2}\hat{\sigma}_z\right) =\exp\left( i\omega_f t \right)\hat{\sigma}_-,
\end{equation}
\begin{equation}\label{pauli operater equation 3}
\exp\left(-\frac{i\omega_f t}{2}\hat{\sigma}_z\right) (\hat{\sigma}_z) \exp\left(\frac{i\omega_f t}{2}\hat{\sigma}_z\right) =\hat{\sigma}_z,
\end{equation}
all relevant terms of the Hamiltonian can be calculated using Eqs. \ref{pauli operater equation 1}-\ref{pauli operater equation 3}.

Their free terms are given by:
\begin{equation}\label{free terms ABC}
\hat{U}(t)\left(\frac{1}{2} \hat{\sigma}^X_z \right) \hat{U}^{\dag}(t)= \frac{1}{2} \hat{\sigma}^X_z,
\end{equation}
where $X\in\{A,B,C\}$; the internal coupling term:
\begin{equation}\label{internal Coupling term}
\hat{U}(t)\hat{H}^{in}\hat{U}^{\dag}(t)= \hat{H}^{in};
\end{equation}
the driving term:
\begin{equation}\label{driving term}
\hat{U}(t)F\left(e^{-i\omega_ft}\hat{\sigma}_+^A+e^{i\omega_ft}\hat{\sigma}_-^A\right)\hat{U}^{\dag}(t) = F\hat{\sigma}_x^A;
\end{equation}
the coherent term of any two subsystems:
\begin{subequations}\label{the coherent term}
\begin{align}
\hat{U}(t)\left(\hat{\sigma}^X_+\hat{\sigma}^Y_- \right) \hat{U}^{\dag}(t) &= \hat{\sigma}^X_+\hat{\sigma}^Y_- ,\\
\hat{U}(t)\left(\hat{\sigma}^X_-\hat{\sigma}^Y_+ \right) \hat{U}^{\dag}(t) &= \hat{\sigma}^X_-\hat{\sigma}^Y_+,
\end{align}
\end{subequations}
where $X,Y\in\{A,B,C\}$;
the derivative term:
\begin{equation}\label{derivative term}
- i \hat{U}(t)\frac{\partial }{\partial t}\hat{U}^{\dag}(t)=-\frac{\omega_f}{2} \left( \hat{\sigma}_z^A+\hat{\sigma}_z^B+\hat{\sigma}_z^C \right).
\end{equation}
From Eqs. \ref{Uf}-\ref{derivative term}, we obtain Eqs. \ref{H_eff_uncat} and \ref{H_eff_cat}.
\bibliography{references}
\bibliographystyle{apsrev4-2}
\end{document}